\documentclass[prl,twocolumn,showpacs,letterpaper,showpacs,superscriptaddress]{revtex4-2}
\usepackage{graphicx,amsmath,amssymb,amsfonts,latexsym,color,dcolumn,bm,epsfig,subfigure}
\usepackage[plainpages=false,hyperfootnotes=false,colorlinks=false]{hyperref}
\usepackage[normalem]{ulem}
\usepackage{dsfont}
\usepackage{tikz}

\renewcommand{\imath}[0]{\mathrm{i}}
\newcommand{\mathbfh}[1]{\hat{\mathbf{#1}}}
\newcommand{\arctanh}[0]{\text{arctanh}}
\newcommand{\abs}[1]{\left\vert#1\right\vert}

\usepackage{orcidlink}

\begin{document}

\title{Nonequilibrium Casimir-Polder Force: Motion-induced Thermal-like Effect}

\author{D. Reiche\orcidlink{0000-0002-6788-9794}}
\affiliation{Humboldt-Universit\"at zu Berlin, Institut f\"ur Physik, 12489 Berlin, Germany}

\author{B. Beverungen\orcidlink{0000-0002-6701-4269}}
\affiliation{Humboldt-Universit\"at zu Berlin, Institut f\"ur Physik, 12489 Berlin, Germany}

\author{K. Busch\orcidlink{0000-0003-0076-8522}}
\affiliation{Humboldt-Universit\"at zu Berlin, Institut f\"ur Physik, 12489 Berlin, Germany}
\affiliation{Max-Born-Institut, 12489 Berlin, Germany}

\author{F. Intravaia\orcidlink{0000-0001-7993-4698}}
\affiliation{Humboldt-Universit\"at zu Berlin, Institut f\"ur Physik, 12489 Berlin, Germany}

\newcommand{\fran}[1]{{\color{blue}#1}}
\newcommand{\dan}[1]{{\color{red}#1}}

\begin{abstract} 
The Casimir-Polder force is analyzed when an atom is moving at a constant velocity relative to a collection of translationally invariant macroscopic bodies with generic shapes and compositions. 
The interaction is described within an approach that accurately treats the atom-field coupling and accounts for the backaction from the environment onto the moving particle. Previously overlooked aspects are uncovered and linked to the nonequilibrium and nonconservative nature of the interaction. Specifically, we examine a behavior that can be understood by characterizing the underlying physical processes in terms of a motional-induced effective temperature. This phenomenon shares similarities with the Fulling-Davies-Unruh effect, opening new perspectives for the understanding of nonequilibrium physics at work in the system.
\end{abstract}

\maketitle

The existence of zero-temperature fluctuation-induced forces is arguably one of the most surprising and least intuitive outcomes of quantum theory~\cite{Schwinger78,Milonni94}.
Among these, a prominent example is the Casimir-Polder force~\cite{London30,Casimir48a,Dzyaloshinskii61}, a quantum electrodynamical interaction between a microscopic electrically neutral, nonmagnetic particle placed in close proximity of other electrically neutral, nonmagnetic objects. Closely related to the van der Waals force, it plays a relevant role in many areas of science and especially in modern quantum technologies, such as atom-interferometers \cite{Cronin09,Hornberger12, Alauze18}, atom-chips \cite{Henkel99, Reichel11, Schneeweiss12, Keil16, Wongcharoenbhorn21}, and atom-fiber systems \cite{Vetsch10, Reitz13}. 
In the past, the Casimir-Polder interaction has been prevalently investigated under the assumption that all system components are fixed in position and in thermal equilibrium, including zero temperature as a special case~\cite{Casimir48a,Dzyaloshinskii61,Intravaia11,Buhmann13}. During the last decades, however, a growing interest in nonequilibrium configurations emerged, including systems featuring thermal gradients and mechanical motion~\cite{Pendry97,Shresta03,Gorza01,Gorza06,Henkel02,Volokitin07,Messina11a,Obrecht07,Scheel09,Rizzuto07,Kruger11,Marino14,Dedkov22,Dedkov21,Silveirinha18,Milton20,Klatt21,Brevik22,Deop-Ruano25}.

The description of nonequilibrium systems is challenging and recent investigations have shown that in some cases relevant aspects were previously overlooked~\cite{Reiche22}. 
Motivated by these findings, we analyze here how the Casimir-Polder force is modified when the particle is moving with respect to other bodies. 
A characteristic feature of this system is its intrinsic nonconservative behavior, an aspect that is important for a consistent thermodynamical description of the undergoing physical processes~\cite{Reiche20c}.
In our self-consistent treatment, we dispense with many common approximations. In particular, we take full account of phenomena like the backaction of the electromagnetic environment on the dynamics of the microscopic particle~\cite{Shresta03,Reiche20c}. Our analysis reveals that the physics of the nonequilbrium Casimir-Polder interaction is actually much richer than previously appreciated.

\begin{figure}
   \includegraphics[width=1.01\linewidth]{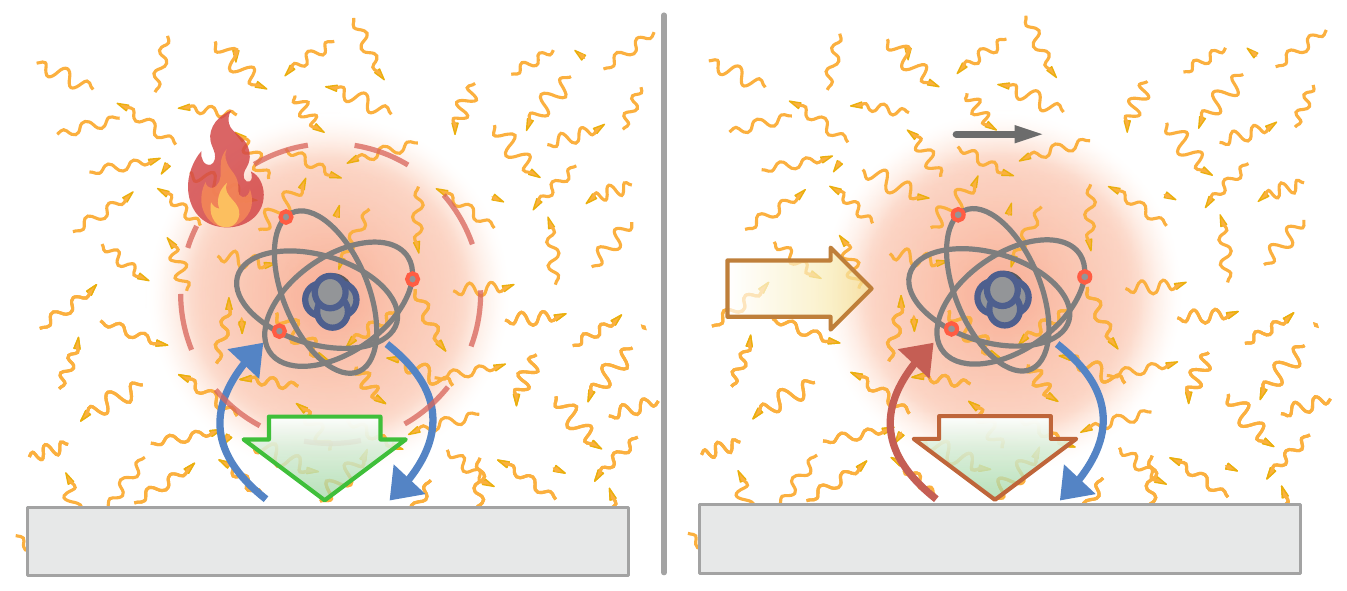}
   \begin{tikzpicture}[remember picture, overlay]
         \node[rotate=0, font=\footnotesize] at (-2.1,4) {\sc Thermal Nonequilibrium};
         \node[rotate=0, font=\footnotesize] at (2.25,4) {\sc Mechanical Nonequilibrium};
        \node[rotate=0, font=\footnotesize] at (-1.1,2.5) {\color{red}$T>0$};
         \node[rotate=0, font=\footnotesize] at (0.8,2.34) {$F_{\rm ext}$};
           \node[rotate=0, font=\footnotesize] at (-2.8,3.3) {$T_{\rm ext}$};
          \node[rotate=0, font=\footnotesize] at (3.35,2.5) {\color{red}$T_{v}>0$};
          \node[rotate=0, font=\footnotesize] at (2.2,3.5) {$v$};
          \node[rotate=0, font=\footnotesize] at (-2.17,1.31) {$\tilde{\mathbf{F}}(T)$};
          \node[rotate=0, font=\footnotesize] at (2.2,1.31) {$\mathbf{F}(v)$};
            \node[rotate=0, font=\footnotesize] at (-3.5,1.1) {\color{blue}$T=0$};
             \node[rotate=0, font=\footnotesize] at (0.85,1.1) {\color{blue}$T=0$};
  \end{tikzpicture}
  \vspace{-0.5cm}
  \caption{Sketch of the analogy discussed here. The Casimir-Polder force on a particle driven at constant velocity $v$ through in the electromagnetic vacuum parallel to an object (right) is similar to the force on a static particle maintained locally in thermal equilibrium at $T>0$ (left). \label{fig:sketch}}
\end{figure}
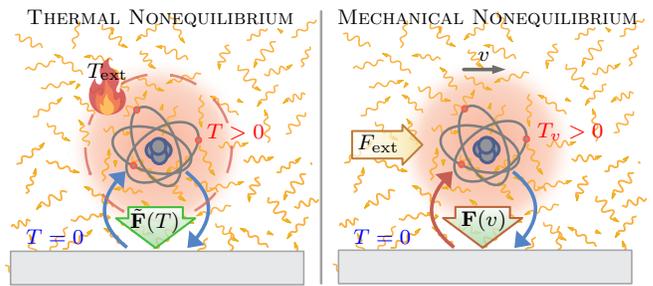

In particular,  we investigate a modification of the interaction, which can be interpreted as arising from a motion-induced increase in the particle's internal temperature. 
This phenomenon shares similarities with the Fulling-Davies-Unruh effect~\cite{Fulling73,Davies75,Unruh76} and previous work~\cite{Intravaia16b} has shown that the emergence of this effective temperature is rooted in the nonconservative nature of the system's light-matter interaction.
Notably, as a result of this behavior, the Casimir Polder force acting on a moving particle resembles the force acting on a particle at rest but maintained at a constant temperature larger than that of its surroundings (see Fig.~\ref{fig:sketch}).

The system we consider consists of a neutral microscopic object moving with non-relativistic speed in vacuum at $T=0$ along 
the translationally invariant direction of an ensemble of macroscopic bodies comprised of passive, linear and reciprocal materials. 
Any magnetic activity is disregarded and the particle's internal quantum dynamics is described by its electric dipole vector operator $\mathbfh{d}$.  
The particle is propelled by an external agent and the total system achieves a nonequlibrium steady state (NESS) with constant velocity when the external force is compensated by quantum friction~\cite{Volokitin07,Reiche22,Milton25}. 
Due to the nonconservative nature of the interaction, we refrain from considering a potential and we define the nonequilibrium Casimir-Polder force as the component of the Lorentz force orthogonal to the velocity~\cite{Intravaia16}. 
We can show that the interaction can be written as 
$\mathbf{F} = \mathbf{F}^{\rm Ds}+ \mathbf{F}^{\rm Th} +\mathbf{F}^{\rm As}$ (see Supplemental Material~\cite{SuppMat}), where
\begin{subequations}
\label{FcpTotal}
\begin{align}
\label{FCPNEqST}
\mathbf{F}^{\rm Ds}
&=\frac{\hbar}{2\pi}\mathrm{Im}\int\limits_0^{\infty}\mathrm{d}\omega
					\int\frac{\mathrm{d}q}{2\pi}
\mathrm{Tr}\left[\underline{\alpha}(\omega_{q}^{-},v)
					\nabla_{\mathbf{R}_a}
					\underline{G}(q,\mathbf{R}_a,\omega)
				\right]~,
\\
\label{FCPNEqTH}
\mathbf{F}^{\rm Th}
&=\int\limits_{0}^{\infty}\mathrm{d}\omega\int\frac{\mathrm{d}q}{2\pi}~
\\ &\times
\mathrm{Tr}\left[\left\{\underline{S}(\omega^{-}_{q},v)-\frac{\hbar}{\pi}\underline{\alpha}_{\Im}(\omega^{-}_{q},v)\right\} \nabla_{\mathbf{R}_{a}}\underline{G}_{\Re}(q, \mathbf{R}_{a}, \omega)\right],
\nonumber\\
\label{FCPNEqAG}
\mathbf{F}^{\rm As}
&=-2\int\limits_{0}^{\infty}\mathrm{d}\omega\,\int\frac{\mathrm{d}q}{2\pi}
\\\nonumber
&\times
\mathrm{Tr}\left[\underline{S}^{\sf T}(-\omega^{-}_{q},v)  \nabla_{\mathbf{R}}\underline{\mathcal G}_{\Im}(q, \mathbf{R},\mathbf{R}_{a}, \omega)\right]_{\vert\mathbf{R}=\mathbf{R}_a}~.
\end{align}
\end{subequations}
In these expressions, $q$ is the component of the wave vector along the direction of motion and $\omega_q^{\pm}=\omega\pm qv$ the Doppler-shifted frequency. The superscript ${\sf T}$ denotes transposition and the vector $\mathbf{R}_{a}$ identifies the particle's constant lateral position, i.e. within the plane orthogonal to the direction of motion.

The nonequilibrium force is given in terms of the trace ($\mathrm{Tr}$) of products of rank-2 tensors related to the physical properties of the system. Specifically, we have defined
\begin{equation}
\label{GIm}
\underline{G}_{\Im}(q,\mathbf{R},\mathbf{R}',\omega)=\frac{\underline{G}(q,\mathbf{R},\mathbf{R}',\omega)-\underline{G}^{\dagger}(q,\mathbf{R}',\mathbf{R},\omega)}{2\imath},
\end{equation}
$\underline{\mathcal G}_{\Im}=[\underline{G}_{\Im}-\underline{G}^{\dag}_{\Im}]/(2\imath)$, and $\underline{G}_{\Re}=\underline{G}-\imath \underline{G}_{\Im}$, where $\underline{G}(q,\mathbf{R},\mathbf{R}',\omega)$ is the Fourier-transformed electromagnetic Green tensor~\cite{Note1}.
For brevity, in Eqs.~\eqref{FcpTotal} and below we retain in the Green tensor only one of the position arguments if $\mathbf{R}=\mathbf{R}'$.
Further, the dipole's power spectrum
\begin{equation}
\label{psd}
\underline{S}(\omega,v)=\int \frac{\mathrm{d}\tau}{2\pi}~\left[\lim_{t\rightarrow\infty}\langle\mathbfh{d}(t)\mathbfh{d}^{\sf T}(t-\tau)\rangle\right] e^{\imath\omega\tau}~,
\end{equation}
describes the quantum statistical properties of the moving particle in the NESS~\cite{Landau80a,Intravaia24}. Finally,  
\begin{equation}
\label{DressPol}
\alpha_{ij}(\omega,v)=\int \mathrm{d}\tau~ \theta(\tau)\left[\lim_{t\rightarrow\infty}\langle \frac{\imath}{\hbar}[\hat{d}_{i}(t),\hat{d}_{j}(t-\tau)]\rangle\right] e^{\imath\omega\tau}~
\end{equation}
are the components of the particle's velocity-dependent dressed-polarizability tensor $\underline{\alpha}$.
In Eq.~\eqref{FCPNEqTH}, the polarizability tensor also appears in the form $\underline{\alpha}_{\Im}=[\underline{\alpha}-\underline{\alpha}^{\dag}]/(2\imath)$~\cite{Intravaia24}. 
In all expressions above, $\mathbfh{d}(t)$ describes the total time evolution of the dipole operator and the expectation value in Eqs.~\eqref{psd} and \eqref{DressPol} is taken over the NESS.

Equations~\eqref{FcpTotal} generalize earlier descriptions of the Casimir-Polder force on a moving particle~\cite{Mahanty80,Ferrell80,Scheel09,Klatt16,Dedkov11,Dedkov21}. Our formulation neither relies on a specific model for the particle's internal dynamics nor on a specific description of the material properties and geometry characterizing its environment.
Within the framework described above, Eqs.~\eqref{FcpTotal} are non-perturbative, non-Markovian and valid beyond the local thermal equilibrium (LTE) approximation~\cite{Intravaia16}. 
The general properties~\cite{Intravaia24} of the quantities entering Eqs.~\eqref{FcpTotal} imply that $\mathbf{F}$ is even in $v$, as one should expect from the system's symmetry. 
Given that the expectation value is taken over the NESS, $\underline{S}$ and $\underline{\alpha}$ depend not only on the particle's velocity $v$ but also on its position $\mathbf{R}_{a}$ (not shown). The latter underscores the contrast to previous work where the force is written as the gradient of a scalar potential~\cite{Mahanty80,Scheel09,Dedkov11,Dedkov21}.

Each of the contributions in Eqs.~\eqref{FcpTotal} can be associated with a specific aspect of the interaction, which is influenced to varying degrees by the nonequilibrium processes within the system. 
The term $\mathbf{F}^{\rm Ds}$ appears as a generalization of the expression for the equilibrium Casimir-Polder force~\cite{CasimirPhysics11,Buhmann13,Intravaia16} and reduces to it for $v\to 0$. The particle's motion manifests itself through the velocity-dependent dressed polarizability and its evaluation at the Doppler-shifted frequency $\omega_{q}^{-}$, which contrary to $\omega$ can become negative (anomalous Doppler-effect). Corrections to the Casimir-Polder force due to the Doppler-shifted frequency have already been considered in the literature~\cite{Mahanty80,Scheel09,Klatt16,Dedkov11,Dedkov21,Volokitin08}. Nonetheless, the dependences on $v$ and $\mathbf{R}_{a}$ of the polarizability and the corresponding physics have received little attention.

Contrary to $\mathbf{F}^{\rm Ds}$, the components $\mathbf{F}^{\rm Th}$ and $\mathbf{F}^{\rm As}$ have no equilibrium counterpart and vanish in the limit $v\to 0$. In other words, they represent pure nonequilibrium modifications of the interaction.
Formally,  $\mathbf{F}^{\rm As}$ shares similarities with the quantum frictional force acting on the particle in the same configuration~\cite{Intravaia19a}. The analysis of $\mathbf{F}^{\rm As}$ is deferred and we focus here on the two remaining components. We want to mention, however, that $\mathbf{F}^{\rm As}$ vanishes for all system's geometries which allow for a mirror symmetry with respect to a plane containing the particle's trajectory. This includes many configurations that have been considered in the literature, which may explain why this term was previously overlooked.

As mentioned above, the expression for $\mathbf{F}=\mathbf{F}^{\rm Ds}+\mathbf{F}^{\rm Th}$ is similar to the expression for the Casimir-Polder force acting on a static particle externally kept locally in thermal equilibrium at $T>0$ by an external agent, e.g. a reservoir, while the remaining system is at zero temperature  (see Fig.~\ref{fig:sketch}). 
Within a perturbative calculation, we obtain in this case
\begin{align}
\label{thermal2}
\tilde{\mathbf{F}}
&\approx \frac{\hbar}{2\pi}\int\limits_{0}^{\infty}\mathrm{d}\xi\, 
\mathrm{Tr}\left[\underline{\tilde\alpha}(\imath \xi)\nabla_{\mathbf{r}_{a}} \underline{G}(\mathbf{r}_{a}, \imath \xi)\right] 
\nonumber\\
&+ \int\limits_{0}^{\infty}\mathrm{d}\omega\;
\mathrm{Tr}\left[\left\{\underline{\tilde S}(\omega)-\frac{\hbar}{\pi}\underline{\tilde \alpha}_{I}(\omega)\right\} \nabla_{\mathbf{r}_{a}}\underline{G}_{R}(\mathbf{r}_{a}, \omega)\right]~,
\end{align}
where $\mathbf{r}_{a}$ is the particle position, $\underline{\tilde \alpha}(\omega)$ is its thermal bare polarizability and $\underline{\tilde S}(\omega)$ is the corresponding dipole's power spectrum (see Supplemental Material~\cite{SuppMat}). Within the LTE approximation they are connected by the fluctuation-dissipation theorem~ \cite{Callen51,Kubo66,Intravaia24}. The subscripts $I$ and $R$ indicate, respectively, the imaginary and real part of the corresponding quantity.
Analogously to $\mathbf{F}^{\rm Ds}$, the first line of Eq.~\eqref{thermal2} recovers the equilibrium force, in this case for $T\to 0$. The term in second line describes the correction due to thermal population of the particle's internal degrees of freedom and its expression is formally analogue to $\mathbf{F}^{\rm Th}$ in Eq.~\eqref{FCPNEqTH}.

To see in detail how this scenario applies to the moving particle, we assume that the dipole operator fulfills~\cite{Intravaia16a}
\begin{equation}
\ddot{\mathbfh{d}}(t)+\omega_a^{2}\mathbfh{d}(t)
=\omega_a^{2} \underline{\alpha}_{0}\cdot \mathbfh{E}(\mathbf{r}_{a}(t),t), \quad 
\underline{\alpha}_{0}=\frac{2\mathbf{d}\mathbf{d}^{\sf T}}{\hbar \omega_a}~,
\label{eqmotion}
\end{equation}
where $ \mathbfh{E}(\mathbf{r},t)$ is the total electric field operator acting on the particle.
This equation provides a simple model for an atom, where $\underline{\alpha}_{0}$ is the static polarizability ($\mathbf{d}$ is a real three-dimensional vector) and $\omega_a$ corresponds to its characteristic transition frequency~\cite{Note2}.
Within this description, $\underline{S}$ and $\underline{\alpha}$ become symmetric tensors and can be calculated explicitly~\cite{Intravaia16a,SuppMat}.
We have
\begin{equation}
\label{model-dressed-pol}
\underline{\alpha}(\omega,v)=
\frac{\underline{\alpha}(\omega)}{1 -\int \frac{\mathrm{d}q}{2 \pi} \mathrm{Tr}\left[ \underline{\alpha}(\omega) \underline{G}(q,\mathbf{R}_{a},\omega^{+}_{q})\right]}~,
\end{equation}
where $\underline{\alpha}(\omega)\equiv\underline{\alpha}_{0}\omega_a^{2}/
[\omega_a^{2} -(\omega+\imath 0^{+})^2]$ is the (causal) bare polarizability for this model~\cite{Note3}.
In addition, according to previous work~\cite{Intravaia16b} we can write
\begin{equation}
\label{Unruh}
\underline{S}(\omega,v)=\frac{\hbar}{\pi}[n_{T_{v}}(\omega)+1]\underline{\alpha}_{I}(\omega,v)~.
\end{equation}
In $n_{T}(\omega)=[e^{\frac{\hbar \omega}{k_{B}T}}-1]^{-1}$ enters the effective temperature
\begin{equation}
\label{effective-temp}
T_{v}= \frac{\hbar v}{2 k_{B} \lambda}
\end{equation}
which describes the particle's quantum statistical properties~\cite{SuppMat}. Consistent with the system being in the NESS~\cite{Reiche20c}, this temperature is constant. 
The length $\lambda$ is connected with the geometric properties of the system, including the particle's position. It weakly depends on the frequency $\omega$, the particle's velocity and the direction of the vector $\mathbf{d}$ (not its strength)~\cite{Intravaia16b}. 

Using Eq.~\eqref{Unruh}, we can rewrite Eq.~\eqref{FCPNEqTH} as
\begin{multline}
\label{thermal}
\mathbf{F}^{\rm Th}=\frac{\hbar}{\pi}\int\limits_{0}^{\infty}\mathrm{d}\omega\int\frac{\mathrm{d}q}{2\pi}~
						 \\\times
						n_{T_{v}}(\omega_{q}^{-})\mathrm{Tr}\left[\underline{\alpha}_{I}(\omega_{q}^{-},v) \nabla_{\mathbf{R}_{a}}\underline{G}_{R}(q; \mathbf{R}_{a}; \omega)\right]~,
\end{multline}
which can be interpreted as a correction arising from the particle's thermal-like excitation. Differently from the static configuration, the thermal gradient is not imposed from the outside but it arises as a consequence of the nonequilibrium processes induced by the particle's motion. 
Physically, the excitation of the particle's internal degrees of freedom can be understood as resulting from the system's quantum mechanical behavior in connection to the anomalous Doppler effect and the quantum Cherenkov effect~\cite{Ginzburg60}: Due to its interaction with a dispersive dissipative environment mediated by the quantum electromagnetic interaction, a part of the particle's kinetic energy is converted into internal energy~\cite{Nezlin76}, effectively heating up its internal degrees of freedom~\cite{Intravaia16b}. 

The similarities between the system with a moving particle and that with a thermal gradient are not only suggestive but also practical for the analysis of $\mathbf{F}^{\rm Th}$. 
For our model, when $T_{v}\gtrsim\hbar\omega_a/k_{\mathrm{B}}\equiv T_{a}$, we can exploit that
$\alpha_{I}$ is strongly peaked around $\pm \omega_a$ and write
\begin{align}
\label{FThHighV}
\mathbf{F}^{\rm Th}
\approx&
\frac{\hbar\omega_a}{2}n_{T_{v}}(\omega_a)\int\limits_{0}^{\infty}\frac{\mathrm{d}\omega}{2\pi}
 \,\frac{\nabla_{\mathbf{R}_{a}}\mathrm{Tr}\left[\underline{\alpha}_{0}\underline{G}_{R}\left(\frac{\omega_a-\omega}{v}; \mathbf{R}_{a}; \omega\right)\right]}{v}
\nonumber\\
+\frac{\hbar\omega_a}{2}&[n_{T_{v}}(\omega_a)+1]\int\limits_{0}^{\infty}\frac{\mathrm{d}\omega}{2\pi}
 \,\frac{\nabla_{\mathbf{R}_{a}}\mathrm{Tr}\left[\underline{\alpha}_{0}\underline{G}_{R}\left(\frac{\omega_a+\omega}{v}; \mathbf{R}_{a}; \omega\right)\right]}{v}.	 				
\end{align}
Notice that, in contrast to the system out of thermal equilibrium~\cite{SuppMat}, due to the behavior of the effective temperature in Eq.~\eqref{effective-temp}, the component $\mathbf{F}^{\rm Th}$ preserves its quantum nature at high values of $T_{v}$. 
After a rapid increase, the expression in Eq.~\eqref{FThHighV} tends to decrease at high velocities~\cite{SuppMat}.
The force in Eq.~\eqref{FThHighV} can be understood as arising from the emission (first term on the r.h.s.) and absorption (second term on the r.h.s.) processes the particle is undergoing during its motion-modified electromagnetic interaction with its surroundings. Their contributions are weighted by the expressions involving the Green tensor which determine their strength and sign.
The second term on the r.h.s. of Eq.~\eqref{FThHighV} is effectively a manifestation of the intrinsic reaction of the moving system, aiming to preserve the system's power balance~\cite{Reiche20c} and the particle's (thermal) stationary state. Only the first term on the r.h.s. of Eq.~\eqref{FThHighV} has an equivalent in the static out-of-equilibrium system~\cite{SuppMat}: 
Since the thermal gradient is kept constant by the external reservoir, no absorption from the environment is needed.
Notice that the terms proportional to $n_{T_{v}}$ does not appear in an approach relying on the LTE approximation~\cite{Dedkov11,Dedkov21}, which partially or even completely neglects the physical processes discussed above. 
For $T_{v}\ll T_{a}$ the integral in Eq.~\eqref{thermal} is dominated by low frequencies. Using a distributional approach~\cite{Estrada02, Estrada02a,SuppMat}, we can write 
\begin{align}
\label{thermalsmallT}
\mathbf{F}^{\rm Th}
&\sim \frac{\pi(k_{B}T_{v})^{2}}{6\hbar}
\mathrm{Tr}\left[\underline{\alpha}'_{I}(0,0) \nabla_{\mathbf{r}_{a}}\underline{G}(\mathbf{r}_{a}; 0)\right]
\\
&+\frac{v^{2}}{4}\frac{\hbar}{\pi}\mathrm{Tr}\left[\underline{\alpha}'_{I}(0,0) \nabla_{\mathbf{R}_{a}}\int\frac{\mathrm{d}q}{2\pi}
						q^{2}\underline{G}(q; \mathbf{R}_{a}; 0)\right]~.
						\nonumber
\end{align}
In this limit, $\mathbf{F}^{\rm Th}$ scales quadratically with the velocity and has two contributions:
The first term on the r.h.s. of Eq.~\eqref{thermalsmallT} is directly connected to motion-induced thermal effect and formally identical to the leading-order correction obtained for the static system with the thermal gradient~\cite{SuppMat}; The second line of Eq.~\eqref{thermalsmallT} appears because $\omega_{p}^{-}$ can become negative~\cite{SuppMat}. This second term has no equivalence in the static system~\cite{SuppMat}. Contrary to the first one, it also appears in a description relying on LTE approximation~\cite{Dedkov11,Dedkov21}. 
Nonetheless, Eq.~\eqref{thermalsmallT} underscores that, in this limit, the approximation inaccurately omits contributions to the force (see left inset of Fig.~\ref{fig:forces} {\bf [b]}).

Further analyses reported in the Supplemental Material~\cite{SuppMat} highlight that, similarly to the equilibrium case~\cite{CasimirPhysics11,Buhmann13,Intravaia16}, $\mathbf{F}^{\rm Ds}$ can be written as an integral over imaginary frequencies~\cite{Dedkov11}
\begin{equation}
\label{FCPNEqST2}
\mathbf{F}^{\rm Ds}
=\frac{\hbar}{2\pi}\int\limits_{0}^{\infty}\mathrm{d}\xi
					\int\frac{\mathrm{d}q}{2\pi}
					\mathrm{Tr}\left[\underline{\alpha}(\imath\xi_{\imath q}^{+},v)
					\nabla_{\mathbf{R}_a}
					\underline{G}(q,\mathbf{R}_a,\imath\xi)
				\right]~,
\end{equation}
where $\imath\xi_{\imath q}^{+}\equiv\imath(\xi+\imath q v)$. 
Owing to the properties of $\underline{\alpha}$ and $\underline{G}$~\cite{Landau80a,Bhatia07,Intravaia24} and given that the symmetric part of the Green tensor is even in $q$, we can show that although $\imath\xi_{\imath q}^{+}$ is complex, after the integration over the wave vector, the integrand is real. 
In the Supplemental Material~\cite{SuppMat}, an analysis of Eq.~\eqref{FCPNEqST2} demonstrates that, due to $\underline{\alpha}(\imath\xi_{\imath q}^{+},v)$, the magnitude of $\mathbf{F}^{\rm Ds}$ is increased relative to the force at rest by a correction $\propto v^2$ at low velocities, but it decreases at high velocities.
This gives rise to non-monotonic behavior as a function of the particle's velocity with the appearance of a maximum.

We have evaluated the previous expressions for a particle that is propelled in vacuum at constant height $z_{a}>0$ within the near field of a planar surface that separates vacuum ($z \ge 0$) from a metallic halfspace ($z \le 0$). We also assume, for simplicity, that $\mathbf{d}$ is orthogonal to the surface. For this configuration $\mathbf{F}^{\rm As}=0$ and the Green tensor is known analytically~\cite{Wylie85,Intravaia16a}. 
The Casimir-Polder force and its two contributions are orthogonal to the surface and the behavior of the magnitudes $F^{\rm Ds}$ and $F^{\rm Th}$ as well as their sum $F$ are depicted in Fig.~\ref{fig:forces} (full lines). These results confirm the features and the asymptotic expressions (black lines) discussed above. For the considered geometry $\lambda \sim z_{a}$~\cite{Intravaia16b}, so that $T_{v}\approx \hbar v/(2k_{B}z_{a}) \equiv \tilde{T}_{v}$ (see the right inset of Fig.~\ref{fig:forces} {\bf [b]}). All quantities are then represented as a function of $\tilde{T}_{v}/T_{a}$, which gives an estimate for the motion-induced thermal excitation of the particle's internal degrees of freedom.

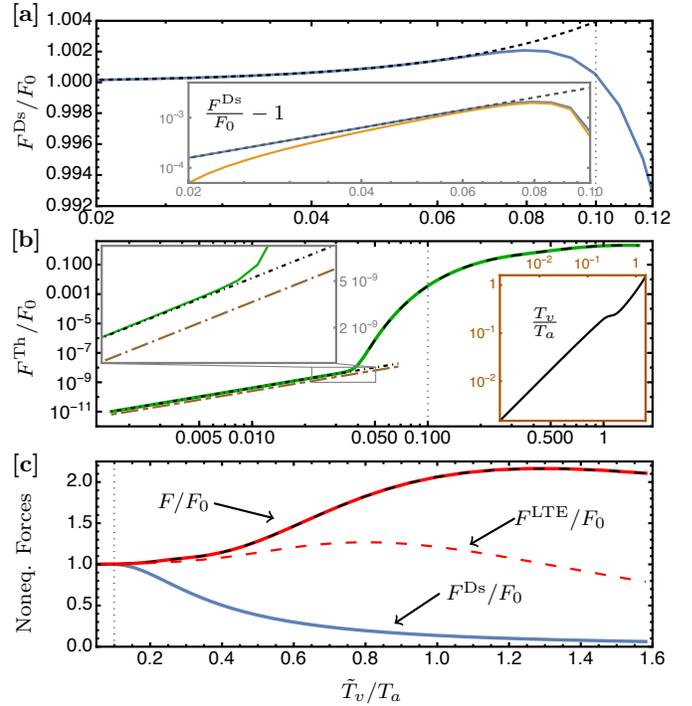
\begin{figure}
   \includegraphics[width=1.01\linewidth]{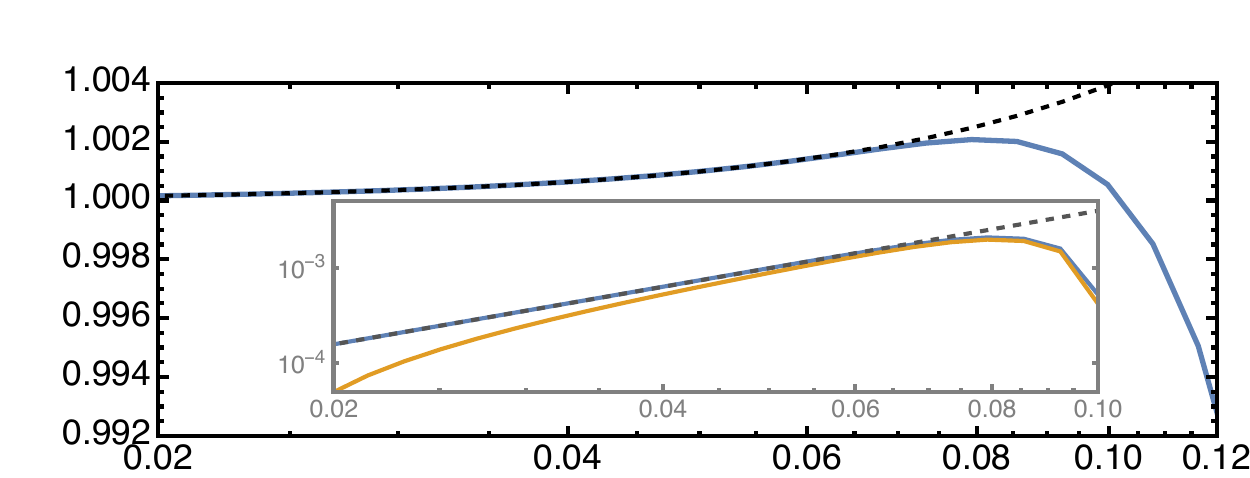}
  \includegraphics[width=1.01\linewidth]{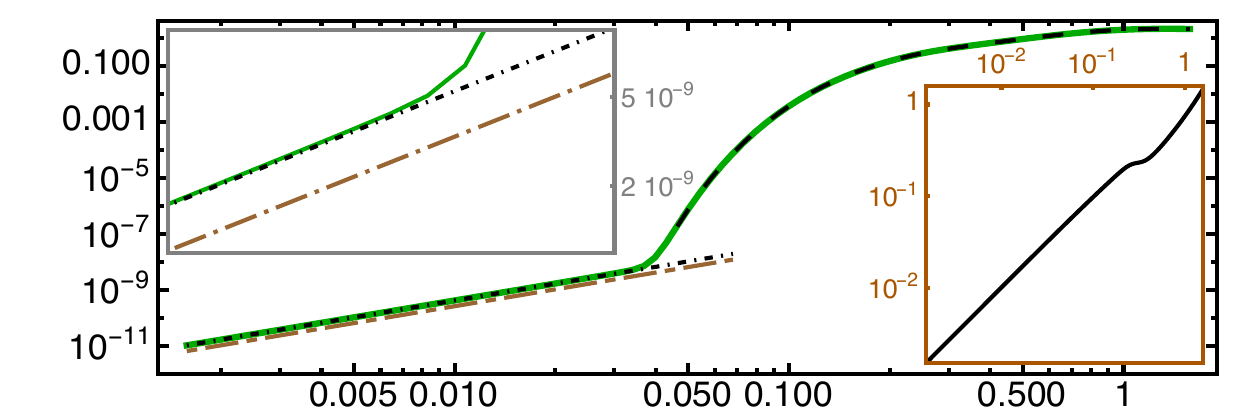}
  \includegraphics[width=1.01\linewidth]{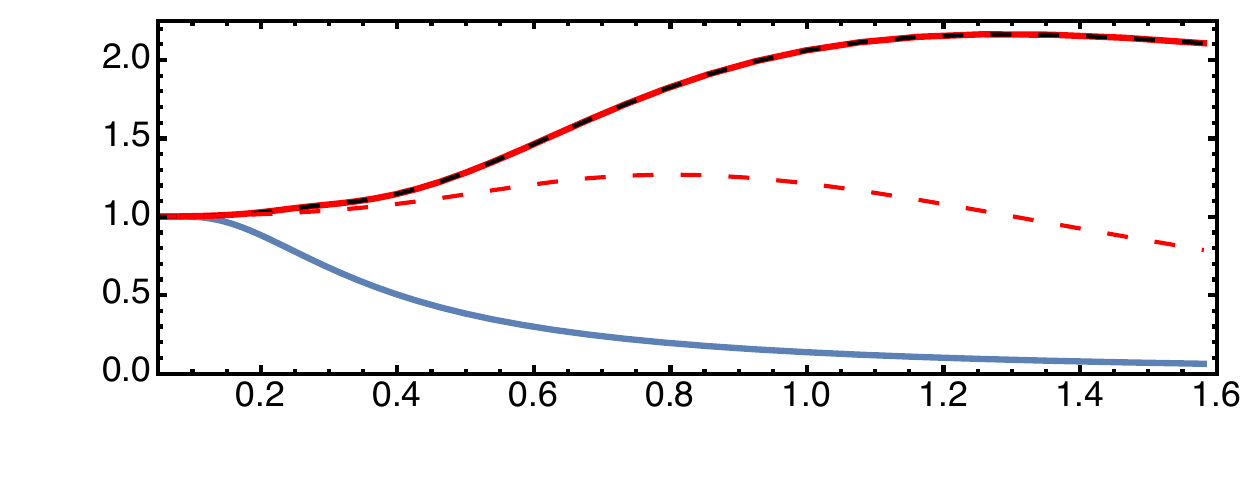}
  \begin{tikzpicture}[remember picture, overlay]
    \node at (-4.1,9.5) {{\bf [a]}};
    \node at (-4.1,6.5) {{\bf [b]}};
     \node at (-4.1,3.5) {{\bf [c]}};
     \node[rotate=90, font=\footnotesize] at (-4.1,8.2) {$F^{\rm Ds}/F_{0}$};
      \node[rotate=90, font=\footnotesize] at (-4.1,5.2) {$F^{\rm Th}/F_{0}$};
       \node[rotate=90, font=\footnotesize] at (-4.1,2.2) {Noneq. Forces};
        \node[font=\footnotesize] at (0.5,0.5) {$\tilde{T}_{v}/T_{a}$};
           \node[font=\footnotesize] at (2.8,5.4) {$\frac{T_{v}}{T_{a}}$};
            \node[font=\footnotesize] at (-1.2,8.2) {$\frac{F^{\rm Ds}}{F_{0}}-1$};
              \draw[dotted] (3.477,7) -- (3.477,9.3);
              \draw[dotted] (1.243,4.15) -- (1.243,6.45);
              \draw[dotted] (-2.926,1.17) -- (-2.926,3.47);
             \draw[gray] (-0.3,4.81) rectangle (0.55,4.61);
             \draw[gray] (-3.1,4.86) -- (-0.3,4.81);
              \draw[gray] (0.01,4.86) -- (0.55,4.81);
              \node[font=\footnotesize] at (-2,3) {$F/F_{0}$};
              \draw[->, thick] (-1.6,3) -- (-0.8,2.8);
              \node[font=\footnotesize] at (2,1.8) {$F^{\rm Ds}/F_{0}$};
              \draw[->, thick] (1.4,1.8)-- (0.8,1.4);
              \node[font=\footnotesize] at (3,2.8) {$F^{\rm LTE}/F_{0}$};
              \draw[->, thick] (2.4,2.8)-- (1.8,2.45);
  \end{tikzpicture}
  \vspace{-0.2cm}
  \caption{The nonequilibrium Casimir-Polder force acting on a particle moving in vacuum within the near field of a planar metallic interface. The different components and the total force are considered for a Cesium atom with $\underline{\alpha}_{0}=\alpha_{0}\mathbf{z}\mathbf{z}$ ($\omega_{a}=1.45$ eV and $\alpha_{0}=4 \pi \epsilon_{0}\times 59.45$ {\AA}$^{3}$~\cite{Steck10a}). The metal is described in terms of the local permittivity function given by the Drude model: $\epsilon(\omega)=1-\omega_{p}^{2}/[\omega(\omega+\imath \gamma)]$ with $\omega_{p}=8.39$ eV and $\gamma/\omega_{p}=10^{-2}$ which corresponds to values for gold. The force and its components are normalized by $F_{0}$, i.e the total force at $v=0$. They are represented as a function of $\tilde{T}_{v}/T_{a}$ for $z_{a}\omega_{a}/c=10^{-1.5}$. With these parameters, we obtain $\tilde{T}_{v}/T_{a}\approx 1.6$ for $v/c=10^{-1}$. 
We marked $\tilde{T}_v/T_{a}=10^{-1}$ with a vertical dotted line in all plots for visual reference. 
{\bf [a]}~$F^{\rm Ds}$ (blue curve) tends to $F_{0}$ for $v\to 0$, deviating from this value as $\propto v^{2}/z_{a}^{6}$ in agreement with the corresponding low velocity asymptotic expression (black dashed curve)~\cite{SuppMat}.  It is visible that $F^{\rm Ds}$ exhibits an extremum, becoming less intense than $F_{0}$ at larger velocities.
Inset: $F^{\rm Ds}$ calculated using the bare polarizability (orange) and full dressed polarizability (blue), see Eqs. \eqref{FCPNEqST} and \eqref{model-dressed-pol}. 
  {\bf [b]}~$F^{\rm Th}$ (green curve) scales as $v^{2}/z_{a}^{9}$ at low velocity, see the Supplemental Material~\cite{SuppMat}. 
  The inset highlights that the force is smaller by about a factor two if the thermal-like effects are neglected. In agreement with the behavior in Eq.~\eqref{FThHighV} (black dashed line), $F^{\rm Th}$ suddenly increases in magnitude when $\tilde{T}_{v}/T_{a}\gtrsim 5\times 10^{-2}$ and tends to flatten.
  Right inset: The effective temperature $T_{v}$ in Eq.~\eqref{effective-temp} normalized by $T_{a}$ as a function of $\tilde{T}_{v}/T_{a}$. The curve quantifies the quality of the approximation $T_{v}\approx\tilde{T}_{v}$.
   {\bf [c]}~The total nonequilibrium Casimir-Polder force $F$ (red curve) is dominated by $F^{\rm Th}$ for $\tilde{T}_{v}/T_{a}> 10^{-1}$, where $F^{\rm Ds}$ (blue curve) becomes a subleading contribution. The black dashed curve describes the total force when $F^{\rm Th}$ is described by Eq.~\eqref{FThHighV}.  The total interaction reaches about twice the equilibrium value for $\tilde{T}_{v}/T_{a}\sim 1$. The red dashed curve shows the force in the LTE approximation, i.e. ignoring the terms proportional to $n_{T_{v}}(\omega_a)$ in Eq.~\eqref{FThHighV}.
    \label{fig:forces}
  }
\end{figure}

In summary, we have shown that, when an atom moves in the electromagnetic vacuum close to translationally invariant objects, the Casimir-Polder force is modified in diverse and unexpected ways.
It can be written as the sum of three contributions, with two of them being the direct result of the nonequilibrium dynamics characterizing the interplay of the particle with the material-modified quantum fluctuations of the electromagnetic field. 
Their expressions are independent of specific models of the particle's internal dynamics and are applicable to a large class of materials and geometries. 
They highlight the nonconservative nature of the interaction, thereby distinguishing our approach from those centered on motion-induced modifications of the Casimir-Polder potential and revealing their inherent limitations.
Specifically, we have shown that the force on a moving particle can closely resemble that acting on a static particle maintained at an higher temperature than its surroundings. Indeed, while it is moving, the particle's internal degrees of freedom are ``heated up'', allowing for the definition of an effective constant temperature~\cite{Intravaia16b}. 
This observation aligns with previous investigations of the Fulling-Davies-Unruh effect, which examine the impact on the Casimir-Polder force of accelerated motion on particles moving in vacuum or close to a surface ~\cite{Rizzuto07,Marino14,Hu04a,Moustos17,Lattuca17}. 
Due to the complexity of the system, these investigations largely use perturbation theory and assume idealized setups (e.g. perfect conductors, two-level atoms, etc.). 
In contrast, in our setup, not only does the motion occur at a constant velocity, but realistic aspects of the system, such as dispersion and dissipation, play a crucial role in ensuring the consistency of our description.
In the example of an atom moving above a planar interface we show that the motion-induced thermal-like effect becomes dominants at high velocities. 
The total nonequilibrium force can also attain twice the equilibrium value, a result that holds significance only within our nonperturbative framework. 

Experimentally, high-velocity neutral atoms can be produced using thermal beams~\cite{Grisenti99,Perreault05,Garcion21} or by accelerating ions -- via ion guns~\cite{Kanitz25}, linear~\cite{Wei12}, or circular~\cite{Denker23} accelerators -- and subsequently neutralizing them within a gas-filled charge-exchange cell~\cite{Schuller07,Rousseau07,Kanitz25}. Detections schemes can include different forms of atom interferometry~\cite{Keil16,Jaffe17,Haslinger17,Cronin09,Hornberger12}, beam-deflection~\cite{Sukenik93} and spectroscopy~\cite{Sandoghdar92,Farias20,Laliotis21,Aquino-Carvalho23}.
Geometry engineering~\cite{Durnin22,Reiche20a} and other optimizations can possibly reduce the challenges associated with the measurement of this phenomenon. 
Additionally, we would like to note that, although in our example the particle is attracted towards the surface, the sign of the interaction generally depends on the particle's properties and, through the Green tensor, the geometry of the system and its material composition. Already at equilibrium, for some specific configurations, the force can be repulsive along some directions~\cite{Levin10,Kristensen23}. 
Further, in static configurations, thermal nonequilibrium is known to lead to repulsion~\cite{Failache99,Gorza01,Antezza05}. Possibly, motion-induced thermal nonequilbrium could give rise to a similar effect. However, this investigation requires care and will be considered in further studies. 
Ultimately, the system considered here bridges diverse areas of physics, offering a platform to explore nonequilibrium light-matter interactions. This framework facilitates the study of phenomena ranging from entropy flow and thermodynamic uncertainty relations~\cite{Horowitz20} to the fundamental characteristics of quantum fluctuation amplification~\cite{Nation12,Oelschlager22}.

\paragraph{Acknowledgements}~-- This project is financially supported by BERLIN QUANTUM. An initiative endowed by the Innovation Promotion Fund of the city of Berlin.



\newpage
\cleardoublepage
\newpage
\setcounter{page}{1}
\setcounter{figure}{0}
\setcounter{equation}{0}
\renewcommand{\theequation}{S\arabic{equation}}
\renewcommand{\figurename}{\textbf{Supplementary Figure}}
\renewcommand{\thefigure}{{\bf S\arabic{figure}}}

\begin{center}
\section{Supplemental Material}
\textit{\large\bf Nonequilibrium Casimir-Polder force: 
Motion-induced Thermal-like Effect\vspace{0.2cm}}

\noindent D. Reiche$^{1}$, B. Beverungen$^{1}$, \\
K. Busch$^{1,2}$ and F. Intravaia$^{1}$.
\end{center}
\begin{footnotesize}
\begin{enumerate}
\item[$^{1}$]
Humboldt-Universit\"at zu Berlin, Institut f\"ur Physik, 12489 Berlin, Germany
\item[$^{2}$]
Max-Born-Institut, 12489 Berlin, Germany
\end{enumerate}
\vspace{0.5cm}
\end{footnotesize}

In the following sections, we provide further details supporting the conclusions in the main text.

\subsection{Expression for the \\nonequilibrium Casimir-Polder force}
\label{S:force}
We provide here the main steps leading to the expressions for the nonequilibrium Casimir-Polder force presented in Eqs.~(1) of the main text.  
The starting point is the Lorentz force acting on the moving particle $\mathbf{F}(t)=\lim_{\mathbf{r}\rightarrow\mathbf{r}_a(t)}\sum_{i} \left\langle\hat{d}_i(t)\nabla_{\mathbf{R}}\hat{E}_i(\mathbf{r},t)\right\rangle$, where $\mathbf{r}_a(t)$ describes the particle trajectory, while $\hat{d}_i$ and $\hat{E}_i$ are, respectively, the $i$-component of the electric dipole vector operator and of the total electric field operator. Proceeding as 
described in Ref.~\cite{Intravaia16}, we obtain that in the stationary limit the force can be written as
\begin{widetext}
\begin{align}
	\mathbf{F}=2\mathrm{Re}\int\limits_{0}^{\infty}\mathrm{d}\omega\,\int\frac{\mathrm{d}q}{2\pi}\int \mathrm{d}\nu~
						\frac{\imath}{\pi}\int\limits_{0}^{\infty}\mathrm{d}\tau \,\mathrm{Tr}\left[
						\underline{S}^{\sf T}(\nu,v)  \nabla_{\mathbf{R}}\underline{G}_{\Im}(q; \mathbf{R},\mathbf{R}_{a}; \omega)_{\vert\mathbf{R}=\mathbf{R}_a}\right]e^{-\imath(\omega_{q}^{-}+\nu)\tau}~,
\end{align}
where $\omega_{q}^{\pm}=\omega\pm qv$ is the Doppler-shifted frequency and we used the tensors defined in the main text. Noticing that
\begin{align}
\nabla_{\mathbf{R}}\underline{G}_{\Im}(q; \mathbf{R},\mathbf{R}_{a}; \omega)_{\vert\mathbf{R}=\mathbf{R}_a}
=\frac{1}{2}\nabla_{\mathbf{R}_{a}}\underline{G}_{\Im}(q; \mathbf{R}_{a}; \omega)
+\imath\nabla_{\mathbf{R}}\underline{\mathcal G}_{\Im}(q; \mathbf{R},\mathbf{R}_{a}; \omega)_{\vert\mathbf{R}=\mathbf{R}_a}
\end{align}
and utilizing the Sokhotski–Plemelj theorem, the expression for the force can be rewritten as 
\begin{align}
\label{S:force-origin}
	\mathbf{F}
		&=\int\limits_{0}^{\infty}\mathrm{d}\omega\,\int \mathrm{d}\nu\int \frac{\mathrm{d}q}{2\pi}~
						 \,\frac{1}{\pi}\mathcal{P}\left(\frac{\mathrm{Tr}\left[
						\underline{S}^{\sf T}(\nu, v) \nabla_{\mathbf{R}_{a}}\underline{G}_{\Im}(q; \mathbf{R}_{a}; \omega)\right]	}{\omega_{q}^{-}+\nu}\right)+\mathbf{F}^{\rm As}~,
\end{align}
where $\mathcal{P}(\dots)$ indicate the principal value and the second term on the r.h.s. has the expression given in Eq.~(1c) of the main text. For the first term on the r.h.s. of Eq.~\eqref{S:force-origin} we use that the power spectrum tensor can be written in a form that resembles that of the fluctuation-dissipation theorem
\begin{align}
\label{S:FDTNEq}
	\underline{S}(\omega, v)=
		\frac{\hbar}{\pi}
		\theta(\omega)\underline{\alpha}_{\Im}(\omega, v)+\underline{J}(\omega,v)~,
\end{align}
where $\underline{J}(\omega,v)$ formally describes a correction arising from a NESS possibly far from equilibrium~\cite{Intravaia24}. The tensors $\underline{\alpha}$ and $\underline{G}$ are susceptibilities with specific analytical properties allowing for the application of the Kramers-Kronig relations~\cite{Jackson75,Landau80a}. In addition, $\underline{S}$ and $\underline{J}$ are Hermitian and the latter, as $\underline{\alpha}$ and $\underline{G}$, fulfills the so-called crossing relation relation, i.e. $\underline{J}(\omega,v) = \underline{J}^{*}(-\omega,v)$~\cite{Intravaia24}.
Inserting Eq.~\eqref{S:FDTNEq} in Eq.~\eqref{S:force-origin} and using the previously described general properties, we arrive at
\begin{align}
\label{S:recombination}
\mathbf{F}
				&=\int\limits_{0}^{\infty}\mathrm{d}\omega\int \frac{\mathrm{d}q}{2\pi}\;
						\frac{\hbar}{2\pi}\mathrm{Im}\mathrm{Tr}\left[
						\underline{\alpha}_{v}(\omega_{q}^{-}) \nabla_{\mathbf{R}_{a}}\underline{G}(q; \mathbf{R}_{a}; \omega)\right]
					\\
				&+\frac{1}{2}\int \frac{\mathrm{d}q}{2\pi}\int \mathrm{d}\omega\;\mathrm{Tr}\left[
\left(\frac{\hbar}{\pi}\left\{ \theta(\omega_{q}^{-})-\theta(\omega)\right\}
\underline{\alpha}_{\Im}(\omega_{q}^{-})+\underline{J}(\omega_{q}^{-},v) \right)\nabla_{\mathbf{R}_{a}}\underline{G}_{\Re}(q; \mathbf{R}_{a}; \omega)\right]+\mathbf{F}^{\rm as}
\nonumber
\end{align}
The first term on the r.h.s. of Eq.~\eqref{S:recombination} gives $\mathbf{F}^{\rm Ds}$ in Eq.~(1a). Considering again Eq.~\eqref{S:FDTNEq}, after some algebra, the second term on the r.h.s. of Eq.~\eqref{S:recombination} can be rewritten in the form given in Eq.~(1b) in the main text.
\end{widetext}

\subsection{The particle's internal dynamics}
\label{S:model}

Although the expressions presented in Eqs.~(1) of the main text are model-independent, and some of their features can be inferred from general properties of the quantities involved~\cite{Intravaia24}, a deeper analysis may require some detail about the particle's internal dynamics. Whenever it has been required, in the main text we have relied on a model developed in previous work~\cite{Intravaia16a}. There, it was shown that when the dipole operator follows a dynamics prescribed by
\begin{subequations}
\label{S:ModelDescription}
\begin{equation}
\ddot{\mathbfh{d}}(t)+\omega_a^{2}\mathbfh{d}(t)
=\omega_a^{2} \underline{\alpha}_{0}\cdot \mathbfh{E}(\mathbf{r}_{a}(t),t), \quad 
\underline{\alpha}_{0}=\frac{2\mathbf{d}\mathbf{d}^{\sf T}}{\hbar \omega_a}~,
\label{S:eqmotion}
\end{equation}
where again $\mathbfh{E}(\mathbf{r},t)$ is the total electric field operator acting on the particle, the power spectum in the NESS
is given by
\label{S:model}
\begin{equation}
\label{S:spectrumNESS}
\underline{S}(\omega,v)=\frac{\hbar}{\pi}\int \frac{\mathrm{d}q}{2\pi}\,\theta(\omega^{+}_{q})\;
\underline{\alpha}(\omega,v)
\underline{G}_{I}(q,\mathbf{R}_{a},\omega^{+}_{q})\underline{\alpha}^{*}(\omega,v)~.
\end{equation}
As in the main text, $\underline{\alpha}(\omega,v)$ is the velocity-dependent dressed polarizability tensor, which can be written as
\begin{equation}
\label{S:model-dressed-pol}
\underline{\alpha}(\omega,v)=
\frac{\underline{\alpha}_{0}\omega_{a}^{2}}{\omega_{a}^{2}-\omega^{2} -\omega_{a}^{2}\int \frac{\mathrm{d}q}{2 \pi} \mathrm{Tr}\left[ \underline{\alpha}_{0} \underline{G}(q,\mathbf{R}_{a},\omega^{+}_{q})\right]}~.
\end{equation}
\end{subequations}
For this model, the bare polarizability $\underline{\alpha}(\omega)$ is obtained by neglecting the term with the Green tensor in the denominator of the previous equation and by enforcing causality. 

The Green tensor $\underline{G}$ can be written in terms of its vacuum and scattered part, $\underline{G}_{0}$ and $\underline{G}_{\rm s}$ respectively.
In our nonrelativistic description ($v/c \ll 1$), the Doppler shift effect can be disregarded for the vacuum part, rendering this contribution independent not only from the position but also effectively from velocity. 
Conversely, given that $\underline{G}_{\rm s}(q,\mathbf{R}_{a},\omega)$ vanishes for $|q|\to \infty$ and $\omega\to \infty$, the contribution from the scattering part tends to vanish for large velocities. In this limit $\underline{\alpha}(\omega,v)$ tends to the known expression for the vacuum dressed polarizability, $\underline{\alpha}^{(v)}(\omega)$ (see, for example, Ref.~\cite{Intravaia11a} for more information).

Consistently with the existence of a NESS, causality and stability are required in the expressions related with the previous model~\cite{Intravaia11a,Silveirinha14a,Intravaia22a}. This is equivalent to require that the poles of the dressed polarizability in Eq.~\eqref{S:model-dressed-pol} are always located in the lower-half part of the complex plane. More concretely, we verify that the condition
\begin{align}
\label{S:passivity}
\int \frac{\mathrm{d}q}{2 \pi} \mathrm{Tr}\left[ \underline{\alpha}_{0} \underline{G}_{I}(q,\mathbf{R}_{a},\omega^{+}_{q})\right]>0 \quad \text{for} \quad \omega\ge 0~
\end{align}
is satisfied for the parameters considered in our analysis.

Although for some evaluations we will refer to the model described in this section, some of the aspects discussed below and in the main text are not directly related to the details of Eq.~\eqref{S:spectrumNESS} and \eqref{S:model-dressed-pol}. They are rather connected to the property of $\underline{S}$ and $\underline{\alpha}$ being symmetric tensors and, as such, are more general.

\subsection{The Casimir-Polder force for an atom in a (motion-induced) thermal state}

As mentioned in the main text, the Casimir-Polder  force $\mathbf{F}=\mathbf{F}^{\rm Ds}+\mathbf{F}^{\rm Th}$ has a form resembling the expression obtained for a particle at rest subjected to a thermal gradient  (see Fig.~1 of the main text).
In this analogy, the particle is assumed to be locally in thermal equilibrium at temperature $T$ close to an object which, in turn, is assumed to be in equilibrium with the radiation at the temperature $T=0$.
The configuration is assumed to be stationary, which requires the existence of an external agent, e.g. a reservoir, preserving the particle's temperature.
 Practically, it is assumed that the probability of finding the particle in one of its internal states is distributed according to thermal statistics. Within perturbation theory, the state of the system is then described by $\hat{\rho}\approx \hat{\rho}_{\rm p}\otimes\hat{\rho}_{\rm env}$, where $\hat{\rho}_{\rm env}$ describes the ground state of the particle's electromagnetic environment, while $\hat{\rho}_{\rm p}=e^{-\frac{\hat{H}_{\rm p}}{k_{B}T}}/Z_{\rm p}$ is the Gibbs state describing the particle's state. 
The operator $\hat{H}_{\rm p}$ is the Hamiltonian of the isolated particle and $Z_{\rm p}$ is the corresponding partition function.

Independently of the specific form of $\hat{H}_{\rm p}$, a perturbative calculation similar to that presented in Ref.~\cite{Intravaia11} gives that the out-of-thermal-equilibrium Casimir-Polder force can be written as
\begin{align}
\label{S:thermal2}
\tilde{\mathbf{F}}
&\approx \frac{\hbar}{2\pi}\int\limits_{0}^{\infty}\mathrm{d}\xi\, 
\mathrm{Tr}\left[\underline{\tilde\alpha}(\imath \xi)\nabla_{\mathbf{r}_{a}} \underline{G}(\mathbf{r}_{a}, \imath \xi)\right] 
\nonumber\\
&+ \int\limits_{0}^{\infty}\mathrm{d}\omega\;
\mathrm{Tr}\left[\left\{\underline{\tilde S}(\omega)-\frac{\hbar}{\pi}\underline{\tilde \alpha}_{I}(\omega)\right\} \nabla_{\mathbf{r}_{a}}\underline{G}_{R}(\mathbf{r}_{a}, \omega)\right]~,
\end{align}
where $\mathbf{r}_{a}$ is the position of the particle in three-dimensional space. In the previous expression we have defined the thermal bare polarizability
\begin{subequations}
\label{S:perturbativeExpressions}
\begin{equation}
\tilde{\alpha}_{ij}(\omega)=\int \text{d}\tau~ \theta(\tau)\langle \frac{\imath}{\hbar}[\hat{d}_{0,i}(t),\hat{d}_{0,j}(t-\tau)]\rangle e^{\imath\omega\tau}~
\end{equation}
and the corresponding dipole's power spectrum
 \begin{equation}
\underline{\tilde S}(\omega)=\int \frac{\text{d}\tau}{2\pi}~\langle\mathbfh{d}_{0}(t)\mathbfh{d}_{0}^{\sf T}(t-\tau)\rangle e^{\imath\omega\tau}~.
\end{equation}
\end{subequations}
Differently from the expressions reported in the main text for the moving particle, Eqs.~\eqref{S:perturbativeExpressions} only involve the free dipole operator evolution $\mathbfh{d}_{0}(t)$ prescribed by the Hamiltonian $\hat{H}_{\rm p}$ and not its full dynamics. In addition, in Eqs.~\eqref{S:perturbativeExpressions}, the average is taken only over $\hat{\rho}_{\rm p}$ and not over the density matrix corresponding to the NESS. 

In the limit $T\to 0$, the first line of Eq.~\eqref{S:thermal2} recovers the corresponding perturbative expression for the zero-temperature equilibrium Casimir-Polder interaction~\cite{CasimirPhysics11}. The second term on the r.h.s. of Eq.~\eqref{S:thermal2} describes instead the correction to the force due to thermal population of the particle's internal degrees of freedom. Within the local thermal equilibrium approximation, the fluctuation-dissipation theorem~ \cite{Callen51,Kubo66,Intravaia24} allows us to write
\begin{equation}
\label{S:FDT}
\underline{\tilde S}(\omega)-\frac{\hbar}{\pi}\underline{\tilde \alpha}_{I}(\omega)=\frac{\hbar}{\pi}n_{T}(\omega)\underline{\tilde \alpha}_{I}(\omega)~,
\end{equation}
where $n_{T}(\omega)=[e^{\frac{\hbar \omega}{k_{B}T}}-1]^{-1}$. When this expression is inserted in Eq.~\eqref{S:thermal2} it takes a form which resembles Eq.~(10) in the main text.

The Bose-Einstein occupation number, $n_{T}(\omega)$, is large for $\omega \lesssim k_{B}T/\hbar\equiv\omega_{\rm th}$ and then exponentially decreases for larger frequencies. As long as $n_{T}(\omega)$ does not overlap with any resonance of the system, in our case for $T\ll T_{a}=\hbar \omega_{a}/k_{B}$, we can use a result of distribution theory~\cite{Estrada02a,Estrada02} to write that
\begin{align}
\label{S:distribAppr}
n(\omega)
\sim -\frac{1}{2}\frac{\pi^{2}}{6}\left(\frac{2k_{B}T}{\hbar}\right)^{2}\delta'(\omega)-\theta(-\omega)~.
\end{align}
The prime indicates the derivative with respect to the frequency, and $\theta(x)$ is the Heaviside step function.
For the second term on the r.h.s. of Eq.~\eqref{S:thermal2}, given that $\omega\ge 0$, only the first term on the r.h.s. of Eq.~\eqref{S:distribAppr} is relevant, showing that the correction to the force due to the particle's thermal population scales $\propto T^{2}$ at low temperature. The situation is different for the motion-induced thermal behavior represented by Eq.~(10) in the main text: The Doppler-shifted frequency $\omega_{q}^{-}$ can become negative making the contribution of the step function significant.
Using Eq.~\eqref{S:distribAppr} in Eq.~(10), for $\mathbf{F}^{\rm Th}$ one obtains the two contributions reported in Eq.~(12) of the main text.

The detailed expression for the high temperature behavior of $\tilde{\mathbf{F}}$ depends on $\underline{\tilde \alpha}(\omega)$. In the case of the model in Eq.~\eqref{S:eqmotion}, we can write that, for $T\gtrsim T_{a}$, 
\begin{align}
\label{S:thermal3}
\tilde{\mathbf{F}}\approx\frac{\hbar}{2\pi}\mathrm{Im}&\int\limits_{0}^{\infty}\mathrm{d}\omega\, 
\mathrm{Tr}\left[\underline{\tilde \alpha}(\omega) \nabla_{\mathbf{r}_{a}}\underline{G}(\mathbf{r}_{a}; \omega)\right] 
\nonumber\\
+&\frac{\hbar\omega_{a}}{2} n(\omega_{a})
\mathrm{Tr}\left[\underline{\alpha}_{0} \nabla_{\mathbf{r}_{a}}\underline{G}_{R}(\mathbf{r}_{a}; \omega_{a})\right]~,
\end{align}
which is the analogue of Eq.~(11) in the main text (see also Eq.~\eqref{S:FThHighV} below).
Notice that, for $T\gg T_{a}$, we can approximate $n_{T}(\omega_{a})\sim k_{B} T/\hbar\omega_{a}$, showing that the second term on the r.h.s. of Eq.~\eqref{S:thermal3} grows $\propto T$ at large temperature,  becoming classical in nature ($\hbar$ disappears).

In accordance with previous work~\cite{Intravaia16b}, in the expressions given in the main text the effective temperature of the moving particle is defined as
\begin{equation}
T_{v}(\omega)=\frac{\frac{\hbar \omega}{k_{\rm B}}}{\ln\left[\frac{\mathcal{D}(\omega, v)}{\mathcal{D}(-\omega, v)}\right]}
\equiv \frac{\hbar v}{2k_{B}\lambda}~.
\label{S:S-Teff}
\end{equation}
Consistent with the balance of power occurring in the NESS~\cite{Reiche20c}, the temperature is a constant in time. Although the system is out-of-equilibrium due to the mechanical motion, the energy flux in and out the particle keep its internal state stable~\cite{Reiche20c}.
The function $\mathcal{D}(\omega, v)$ is given in terms of the Green tensor
\begin{align}
\mathcal{D}(\omega, v)
&=\int \frac{\mathrm{d}q}{2\pi}\,\theta(\omega_{q}^{+})\;
\mathrm{Im}\mathrm{Tr}\left[\underline{\alpha}_{0}\underline{G}_{\rm s}(q,\mathbf{R}_{a},\omega_{q}^{+})\right]
\nonumber\\
&+\theta(\omega)\frac{\mathrm{Tr}[\underline{\alpha}_{0}]}{6\pi\epsilon_{0}}\left(\frac{\omega}{c}\right)^{3},
\end{align} 
where the second term on the r.h.s. is directly related to the vacuum part, $\underline{G}_{0}$.
For the motion-modified Casimir-Polder force, the term equivalent to the second line of Eq.~\eqref{S:thermal3} is provided in Eq.~(11) of the main text and reproduced below for direct comparison.
\begin{align}
\label{S:FThHighV}
\mathbf{F}^{\rm Th}
\approx&
\frac{\hbar\omega_a}{2}n_{T_{v}}(\omega_a)\int\limits_{0}^{\infty}\frac{\mathrm{d}\omega}{2\pi}
 \,\frac{\nabla_{\mathbf{R}_{a}}\mathrm{Tr}\left[\underline{\alpha}_{0}\underline{G}_{R}\left(\frac{\omega_a-\omega}{v}, \mathbf{R}_{a}, \omega\right)\right]}{v}
\nonumber\\
+\frac{\hbar\omega_a}{2}&[n_{T_{v}}(\omega_a)+1]\int\limits_{0}^{\infty}\frac{\mathrm{d}\omega}{2\pi}
 \,\frac{\nabla_{\mathbf{R}_{a}}\mathrm{Tr}\left[\underline{\alpha}_{0}\underline{G}_{R}\left(\frac{\omega_a+\omega}{v}, \mathbf{R}_{a}, \omega\right)\right]}{v}.	 				
\end{align}
To obtain this expression we used that $n_{T}(-\omega)=-n_{T}(\omega)-1$.
Despite the similarities, we want to mention that Eq.~\eqref{S:FThHighV} has its own specific features when compared to its static out-of-thermal-equilibrium counterpart. While for $T_{v}\gg T_{a}$ the function $n_{T_{v}}(\omega_a)$ linearly scales with $T_{v}$, due to the nature of the effective temperature, the behavior of $\mathbf{F}^{\rm Th}$ in Eq.~\eqref{S:FThHighV} is still quantum. Specifically, at large velocities we can approximate Eq.~\eqref{S:FThHighV} as
\begin{equation}
\label{S:FThHighV2}
\mathbf{F}^{\rm Th}
\approx
\frac{k_{B} T_{v}}{2}\int \frac{\mathrm{d}q}{2\pi}
 \,\nabla_{\mathbf{R}_{a}}\mathrm{Tr}\left[\underline{\alpha}_{0}\underline{G}_{R}\left(q, \mathbf{R}_{a}, \omega_{a}+qv\right)\right].
\end{equation}
Given that  $\nabla_{\mathbf{R}_{a}}\mathrm{Tr}\left[\underline{\alpha}_{0}\underline{G}_{R}\left(q; \mathbf{R}_{a}; \omega \right)\right]\propto \omega^{-2}$ at large frequencies, for $T_{v}\propto v$, we have that $\mathbf{F}^{\rm Th}$ tends to decrease at large velocities. 
Despite this mathematical behavior, the previous consideration must be interpreted within the physical constraints of our nonrelativistic description.

\subsection{The behavior of $\mathbf{F}^{\rm Ds}$}

In order to provide a full description of the nonequilibrium Casimir-Polder interaction, it is relevant to analyze the properties of $\mathbf{F}^{\rm Ds}$.
Causality and stability require that the frequency integrand of Eq.~(1a) must be analytic in the upper-half part of the complex-frequency plane. 
As for the equilibrium interaction~\cite{Intravaia11}, this allows us to perform a Wick rotation to write 
\begin{align}
\label{S:FCPNEqST2}
\mathbf{F}^{\rm Ds}
&=\frac{\hbar}{2\pi}\int\limits_{0}^{\infty}\mathrm{d}\xi
					\int\frac{\mathrm{d}q}{2\pi}
					\mathrm{Tr}\left[\underline{\alpha}(\imath\xi_{\imath q}^{+},v)
					\nabla_{\mathbf{R}_a}
					\underline{G}(q,\mathbf{R}_a,\imath\xi)
				\right]~,
\end{align}
where $\imath\xi_{\imath q}^{\pm}\equiv\imath(\xi\pm\imath q v)$. 
Notice that whenever $\underline{\alpha}(\imath\xi,v)$ is a symmetric tensor, as for the model considered in the main text (see Eqs.~\eqref{S:ModelDescription}), we can write~\cite{Intravaia24}
\begin{equation}
\label{S:alphasymm1}
\underline{\alpha}(\imath\xi_{\imath q}^{+},v)=\frac{2}{\pi}\int\limits_{0}^{\infty}\mathrm{d}\omega \frac{\omega\underline{\alpha}_{I}(\omega,v)}{\omega^{2}+(\xi+\imath q v)^{2} }~.
\end{equation}
Given that the trace in Eq.~\eqref{S:FCPNEqST2} selects the symmetric part of the Green tensor which is even in $q$~\cite{Intravaia24}, only the part of $\underline{\alpha}(\imath\xi_{\imath q}^{+},v)$ which is even in $q$ is relevant in Eq.~\eqref{S:FCPNEqST2}. This means that there we can replace $\underline{\alpha}(\imath\xi_{\imath q}^{+},v)$ with 
\begin{multline}
\underline{\alpha}_{e}(\imath\xi_{\imath q}^{+},v)\equiv 
\frac{\underline{\alpha}(\imath\xi_{\imath q}^{+},v)+\underline{\alpha}(\imath\xi_{\imath q}^{-},v)}{2}
\\=\frac{2}{\pi}
\int\limits_{0}^{\infty}\mathrm{d}\omega\frac{\omega(\omega^{2}+\xi^{2}-q^{2}v^{2})\underline{\alpha}_{I}(\omega,v)}{(\omega^{2}+\xi^{2}-q^{2}v^{2})^{2}+4 (\xi q v)^{2} }
~.
\label{S:alphasymm2}
\end{multline}
Since the denominator of the previous expression is positive for $\xi \in\mathbb{R}$, using $\underline{\alpha}_{e}(\imath\xi_{\imath q}^{+},v)$ in Eq.~\eqref{S:FCPNEqST2} underscores that the integral in Eq.~\eqref{S:FCPNEqST2} is real.
While in general $\underline{\alpha}(\imath\xi,v)$ is real and has elements which are monotonically decreasing functions of $\xi$~\cite{Landau80a,Intravaia24}, if it is symmetric, it is also positive semidefinite for $\xi\ge 0$. The latter changes, however, when $\xi$ is replaced with $\xi_{\imath q}^{+}$. 
Indeed, $\underline{\alpha}_{I}(\omega,v)$ is positive semidefinite and in general peaked around the characteristic frequency $\omega_{a}$. Consequently, $\underline{\alpha}_{e}(\imath\xi_{\imath q}^{+},v)$ is positive semidefinite for all $\xi$ only if $|qv|<\omega_{a}$.
In addition, according to the Loewner order~\cite{Bhatia07}, we have that $\underline{\alpha}_{e}(\imath\xi_{\imath q}^{+},v)>\underline{\alpha}(\imath\xi,v)$ for $0\le\xi\lesssim\sqrt{(\omega_{a}^{2}-q^{2}v^{2})/3}$, while the opposite ordering occurs for larger $\xi$.  Conversely, when $|qv|>\omega_{a} $ then $\underline{\alpha}_{e}(\imath\xi_{\imath q}^{+},v)$ is negative semidefinite for $0\le\xi\lesssim\sqrt{q^{2}v^{2}-\omega_{a}^{2}}$, becoming again positive semidefinite for larger $\xi$. A similar argument can be made for the denominator of Eq.~\eqref{S:model-dressed-pol}, indicating that, for the specific model in Eq.~\eqref{S:eqmotion}, independently of the detailed composition and geometry of the electromagnetic environment, the contribution of $\underline{\alpha}(\imath\xi,v)$ to Eq.~\eqref{S:FCPNEqST2} is enhanced for small $v$ but diminished at large velocities.
As pointed out in the main text, the properties of  $\underline{\alpha}_{e}(\imath\xi_{\imath q}^{+},v)$ in conjunction with the integration over $q$, give rise to a nonmonotonic behavior as a function of the particle's velocity, with an initial enhancement  followed by a reduction of $\mathbf{F}^{\rm Ds}$ with respect to its equilibrium value and to the appearance of at least one extremum (see Fig.~1{\bf [a]} in the main text). 

The impact of the polarizability's behavior on the Casimir Polder force can be better understood by investigating in detail the low and high velocity behavior of $\mathbf{F}^{\rm Ds}$. Using the expression for the equilibrium force,
\begin{equation}
\label{S:F0SM}
\mathbf{F}_{0}=\frac{\hbar}{2\pi}\int\limits_{0}^{\infty}\mathrm{d}\xi\;
\mathrm{Tr}\left[\underline{\alpha}(\imath \xi,0) \nabla_{\mathbf{r}_{a}}\underline{G}(\mathbf{r}_{a}; \imath \xi)\right]
\equiv\mathbf{F}^{\rm Ds}_{\vert v=0}~,
\end{equation}
we can write $\mathbf{F}^{\rm Ds}=\mathbf{F}_{0}+\Delta\mathbf{F}^{\rm Ds}$, where $\Delta\mathbf{F}^{\rm Ds}$ is the motion-induce correction. At second order in $v$ for the model in Eqs.~\eqref{S:ModelDescription} we can write  
\begin{align}
\label{S:DeltaFDs}
\Delta\mathbf{F}^{\rm Ds}
&\sim \frac{v^{2}}{2}\frac{\hbar}{2\pi}\left(
\int\limits_{0}^{\infty}\mathrm{d}\xi\;
\mathrm{Tr}\left[\underline{\alpha}(\imath \xi,0)
\nabla_{\mathbf{r}_{a}}\underline{G}(\mathbf{r}_{a}, \imath \xi)\right]\right.
\nonumber\\
&\hspace{1.8cm}\times\mathrm{Tr}\left[\underline{\alpha}(\imath \xi,0) \int \frac{\mathrm{d}q}{2 \pi} q^{2} \underline{G}''(q,\mathbf{R}_{a},\imath \xi)\right]
\nonumber\\
&+\left.
\int\limits_{0}^{\infty}\mathrm{d}\xi\;\mathrm{Tr}\left[\underline{\alpha}''(\imath \xi,0)\int\frac{\mathrm{d}q}{2\pi}q^{2} \nabla_{\mathbf{R}_{a}}\underline{G}(q, \mathbf{R}_{a}, \imath \xi)\right]\right).
\end{align}
In Eq.~\eqref{S:DeltaFDs}, the first contribution in the parentheses stems from the intrinsic velocity dependence of the dressed polarizability, while the second originates from the dependence of the dressed polarizability on the Doppler-shifted frequency.
At leading order in $\underline{\alpha}_{0}$ only the second term is relevant and we can write
\begin{align}
\label{S:correction2ndOrder}
\Delta\mathbf{F}^{\rm Ds}
&\sim \frac{v^{2}}{2}\frac{\hbar}{2\pi}
\int\limits_{0}^{\infty}\mathrm{d}\xi\;\mathrm{Tr}\left[\underline{\alpha}''(\imath \xi)\int\frac{\mathrm{d}q}{2\pi}q^{2} \nabla_{\mathbf{R}_{a}}\underline{G}(q, \mathbf{R}_{a}, \imath \xi)\right],
\end{align}
where $\underline{\alpha}''(\omega)$ is the second derivative in $\omega$ of the bare polarizability. This implies $\underline{\alpha}''(\imath \xi)=-d^{2}\underline{\alpha}(\imath \xi)/\mathrm{d}\xi^{2}$, which is positive for $0<\xi\le \omega_{a}/\sqrt{3}$ and negative otherwise.

The behavior of $\mathbf{F}^{\rm Ds}$ at large $v$ is more directly related to the Green tensor. To address this aspect, it is convenient to write Eq.~\eqref{S:FCPNEqST2} in an equivalent form. Using Eq.~\eqref{S:alphasymm2}, after performing a partial integration in the variable $q$, this leads to
\begin{align}
\label{S:FCPNEqST3}
\mathbf{F}^{\rm Ds}
=-\hbar\int\limits_{0}^{\infty}\frac{\mathrm{d}\xi}{\pi}\int\limits_{0}^{\infty}&\frac{\mathrm{d}\omega}{\pi}
					\int\limits_{0}^{\infty}\frac{\mathrm{d}q}{\pi}\;
				\frac{\arctanh \left[\frac{2\frac{\omega}{v}\; q}{\frac{\xi^2+\omega^{2}}{v^{2}} +q^2}\right]}{v}
				\nonumber\\
				& \times	
					\mathrm{Tr}\left[\underline{\alpha}_{I}(\omega,v)
					\nabla_{\mathbf{R}_a}
					\underline{G}_{q}(q,\mathbf{R}_a,\imath\xi)
				\right]~,
\end{align}
where the subscript $q$ indicates the derivative with respect to this variable. 
Since the inverse hyperbolic function is positive for $q>0$ and $\underline{\alpha}_{I}(\omega,v)$ is positive semidefinite for $\omega>0$, the sign of the integrand in Eq.~\eqref{S:FCPNEqST3} and therefore the behavior of the force, is connected with $\nabla_{\mathbf{R}_a}\underline{G}_{q}(q,\mathbf{R}_a,\imath\xi)$ for $q>0$. 
More specifically, given that the elements of the Green tensor monotonically decrease to zero as a functions of $\xi$~\cite{Landau80a,Intravaia24}, the sign of the integrand is related to the monotonicity (or lack thereof) in the variable $q$.
Given that the function in the second line in Eq.~\eqref{S:FCPNEqST3} is odd in $q$ and vanishes for $q\to 0$ as well as for $q\to\infty$, the largest contribution to the integral over this variable arises from finite nonzero values. 
In the limit $v\to 0$ this allows to recover expressions for $\mathbf{F}_{0}$ and $\Delta\mathbf{F}^{\rm Ds}$ equivalent to Eqs.~\eqref{S:F0SM} and \eqref{S:DeltaFDs}. Conversely, at large velocities, given that $\underline{\alpha}_{I}(\omega,v)$ and the Green tensor effectively
limit the values of $\omega$ and $\xi$, we can conclude from Eq.~\eqref{S:FCPNEqST3} that $\mathbf{F}^{\rm Ds}$ vanishes as $\propto v^{-2}$ or faster. 
More precise considerations require a detailed knowledge of the Green tensor. Once again, the physical relevance of these mathematical results should, however, be vetted by the limits of our nonrelativistic framework.

\subsection{The special case of a plane interface}

The previous expressions are still rather general since the atom’s electromagnetic environment is not specified. 
We specify here the system by considering a particle moving in the near field of a planar surface that separates vacuum ($z> 0$) from a metallic half space ($z \le 0$). The system's symmetry requires that the total force and its components are all oriented orthogonally to the interface, which allows us to focus only on their magnitudes.

As mentioned above, the Green tensor is the sum of a vacuum and a scattered part. For our evaluation, the relevant part of the former is 
\begin{equation}
\label{S:Gdef}
\mathrm{Im}\left[\underline{G}_{0}(\omega)\right]=\imath\frac{\left(\frac{\omega}{c}\right)^{3}}{6\pi\epsilon_{0}} \underline{1},
\end{equation}
wich is effectively a scalar ($\underline{1}$ is the identity matrix) due to the isotropy and homogeneity of vacuum. The real part of $\underline{G}_{0}$, associated with the vacuum Lamb-shift~\cite{Milonni94},  is absorbed in the particle's transition frequency.
As mentioned above, due to our nonrelativistic framework we can neglect the Doppler-shift in this term. This allows us perform the $q$ integration to arrive at Eq.~\eqref{S:Gdef}.

The expression of the scattered part of the Green tensor for a planar structure is known exactly~\cite{Wylie85,Intravaia16a}. For simplicity, we consider here only its near-field limit~\cite{Intravaia16,Intravaia19a} which provides the dominant contribution to the interaction. For a vacuum material interface, we have
\begin{gather}
\underline{G}_{s}(q,z_{a}, \omega)\sim \frac{r(\omega)}{2\epsilon_{0}} 
			 \int \frac{dp}{2\pi} \underline{\Pi} \;k e^{-2k z_{a}}~,
\\\nonumber
\text{where}\quad \underline{\Pi}
=
\begin{pmatrix}
\frac{q^{2}}{k^{2}}&\frac{pq}{k^{2}}&-\imath \frac{q}{k}\\
\frac{pq}{k^{2}}&\frac{p^{2}}{k^{2}}&-\imath \frac{p}{k}\\
\imath \frac{q}{k}&\imath \frac{p}{k}&1
\end{pmatrix},			 
\end{gather}
$k=\sqrt{q^{2}+p^{2}}$, $\epsilon_{0}$ is the vacuum permittivity and $r(\omega)\in \mathbb{C}$ is the p-polarized reflection coefficient at the interface with a (spatially local) material. 
Moreover, we choose for simplicity a dipole oriented along the $z$-axis, $\mathbf{d}=d\mathbf{z}$ so that
\begin{equation}
\underline{\alpha}_{0}=\alpha_{0}\mathbf{z}\mathbf{z}^{\sf T}\quad\text{with}\quad \alpha_{0}=\frac{2d^{2}}{\hbar\omega_{a}}~.
\end{equation}
This also restricts the calculation to the $zz$-element of the Green tensor, which can be also written as
\begin{equation}
[\underline{G}_{s}(q,z_{a}, \omega)]_{zz}\sim
\frac{r(\omega)}{2\epsilon_{0}}\frac{q^{2}}{2\pi} [K_{0}(2\abs{q}z_{a})+K_{2}(2\abs{q}z_{a})]~,
\end{equation}
where $K_{n}(x)$ are the modified Bessel functions of the second kind and order $n$.

Given the Green tensor, we can evaluate the asymptotic expressions presented in the main text and those derived above. For example, we can show~\cite{Intravaia16}
\begin{equation}
T_{v}(\omega)\xrightarrow{\omega\to 0}
\frac{3}{\pi}\frac{\hbar v}{2k_{B}z_{a}} \text{ and } T_{v}(\omega)\xrightarrow{\omega\to \infty}\frac{\hbar v}{2k_{B}z_{a}}.
\end{equation}
The first limit can be used for evaluating the low temperature (and low velocity) limit of $F^{\rm th}$  (Eq.~(9) in the main text) for the geometry considered above. We have
\begin{align}
F^{\rm Th}
&\approx -\left[\left(\frac{3}{\pi}\right)^{2}+\frac{15}{\pi^{2} }\right]\frac{\hbar}{\pi}
\left(\frac{\alpha_{0}}{2\epsilon_{0}} \right)^{2} 
\frac{v^{2}}{ (2z_{a})^9}
r'_{I}(0) 
 r(0)~.
\end{align}
The first term in the square brackets is due the motion induced thermal excitation of the atom (first line of Eq.~(9)) while the second term is related to the anomalous Doppler effect (second line of Eq.~(9)). 

Considering Eq.~\eqref{S:correction2ndOrder} for the planar geometry, at the second order in the velocity we can write 
\begin{equation}
\label{S:asymptFDs}
F^{\rm Ds}\sim F_{0}\left(1+\frac{F_{B}}{F_{0}}\frac{5}{4}\frac{v^{2}}{z_{a}^2}\frac{\int\limits_{0}^{\infty}\mathrm{d}\xi\;\mathrm{Tr}\left[\underline{\alpha}''(\imath \xi) \right]r(\imath \xi)}{\int\limits_{0}^{\infty}\mathrm{d}\xi\;
\mathrm{Tr}\left[\underline{\alpha}(\imath \xi)\right] r(\imath\xi)}\right)~,
\end{equation}
where $F_{B}$ is the equilibrium Casimir-Polder force calculated using the bare polarizability. The ratio $F_{B}/F_{0}$ gives an estimate of the impact of the electromagnetic dressing on the equilibrium Casimir-Polder interaction. Since for an atom in front of a plane $F_{B}\propto z_{a}^{-4}$, this correction corresponds to a deviation of $F^{\rm Ds}$ from the equilibrium value that scales as $\propto v^{2}/z_{a}^{6}$~\cite{Ferrell80,Dedkov11}.

The previous expressions do not depend on the form of $r(\omega)$. For the numeric evaluation
presented in Fig.~1 of the main text we use
\begin{equation}
r(\omega)=\frac{\epsilon(\omega)-1}{\epsilon(\omega)+1}, \quad \epsilon(\omega)=1-\frac{\omega_{p}^{2}}{\omega(\omega+\imath \gamma)},
\end{equation}
where $\omega_{p}$ is the plasma frequency and $\gamma$ the material dissipation rate. Their values are given in the caption of Fig.~1.

\end{document}